\pdfoutput=1
\documentclass[a4paper,12pt]{iopart}
\usepackage{graphicx}

\begin{document}

\title[High Energy Neutrino Telescopes]{High Energy Neutrino Telescopes}

\author{K D Hoffman}

\address{Department of Physics, University of Maryland, College Park, MD 20742 USA}
\ead{kara@umd.edu}
\begin{abstract}
This paper presents a review of the history, motivation and current status of high energy neutrino telescopes. Many years after these detectors were first conceived, the operation of kilometer-cubed scale detectors is finally on the horizon at both the South Pole and in the Mediterranean Sea. These new detectors will perhaps provide us the first view of high energy astrophysical objects with a new messenger particle and provide us with our first real glimpse of the distant universe at energies above those accessible by gamma-ray instruments. Some of the topics that can be addressed by these new instruments include the origin of cosmic rays, the nature of dark matter, and the mechanisms at work in high energy astrophysical objects such as gamma-ray bursts, active galactic nuclei, pulsar wind nebula and supernova remnants.
\end{abstract}


\section{Introduction}

Neutrino astronomy was born with the first detection of low energy neutrinos from an astrophysical source, the Sun, by the pioneering work of Ray Davis and John Bahcall. Using a  100,000 gallon tank of perchloroethylene in the Homestake mine, Ray Davis and his group painstakingly extracted and counted Argon molecules created through inverse beta decay by solar neutrinos captured on Chlorine atoms as they traversed the tank.   The flux of neutrinos inferred from this careful measurement was a mere 1/3 of that calculated by Bahcall for models of solar fusion~\cite{davis, bahcall}. The disagreement between the measurements of Davis and the predictions of Bahcall led to what was known as the Òsolar neutrino problemÓ. 

In the years since, ongoing inquiries into the deficit of solar neutrinos by the Super-Kamiokande and SNO observatories,  along with radiochemical measurements using gallium, have yielded surprising revelations about the very nature of the messengers themselves~\cite{SK, SNO, sage, gallex}.  These collaborations have established that solar neutrinos undergo flavor oscillations, explaining the discrepancy first suggested by Davis' observations decades before, and showing that neutrinos are not the massless  particles once thought~\cite{sk-atm}.  


The first observation of neutrinos from a source outside of our solar system came quite fortuitously on February 23, 1987 at 7:35 UT when a burst of neutrinos from the collapse of a blue supergiant in the Large Magellanic Cloud, was simultaneously detected by the IMB~\cite{IMB} and Kamiokande-II~\cite{ka1987} detectors.  A total of 19 neutrinos were observed within a time interval of about 6 seconds. 
It was sighted in the optical three hours later. 
 This event, which became known as Supernova 1987A, was the nearest observed supernova explosion to earth in nearly 400 years and therefore provided a rare opportunity, not only to confirm models of supernova formation, but to constrain basic properties of neutrinos, including their masses and lifetimes.  This again established the unique potential of neutrino telescopes, both as astrophysical observatories and as long baseline neutrino detectors.

By the time Supernova 1987A was observed, the idea of using neutrinos as messengers to study the universe at the highest energies had already been conceived.  In the ensuing years, the technology and methods required to build an array sensitive enough for extragalactic observations were developed through the construction and operation of a number of arrays on an intermediate scale.
The first detectors on the scale of a cubic kilometer, which are large enough to be sensitive to the expected flux of astrophysical neutrinos, are now nearing completion and are coming into operation. These new neutrino telescopes are truly discovery instruments. They hold the promise of opening a new window on the high energy universe by extending the observable energy range for distant and optically opaque objects and by using a new particle to complement the many observations using gamma rays. 

This paper begins by explaining the motivation to use neutrinos as astrophysical messengers at high energies, the connection of the expected neutrino flux to the known cosmic ray flux, and why detectors on the scale of a kilometer are necessary to reach a sensitivity of astrophysical interest. In section two, the development of the techniques used in neutrino astronomy and the performance requirements common to all detectors of this type are described followed by the status of the major neutrino telescopes now in operation or under construction. Section three presents the results from current experiments along with the expected sensitivity for the new detectors that are coming on line. 

\subsection{Why neutrinos and why a kilometer cubed?}

One of the most compelling arguments for building large scale neutrino observatories is to shed light on the mystery of cosmic rays.
Since their discovery almost 100 years ago, the origin of cosmic rays has presented a puzzle. The energy spectrum and flux of charged particle cosmic rays bombarding the Earth has now been extensively measured, and the range of the observed energies is truly astounding, with a handful of events recorded with an energy in excess of $10^{20}$ eV (Figure 1 right). These measurements have established the fact that there are cosmic charged particle accelerators. Yet, to this day, no unambiguous source of hadronic acceleration has been found. 

\begin{figure}[htbp] 
  \includegraphics[width=3in]{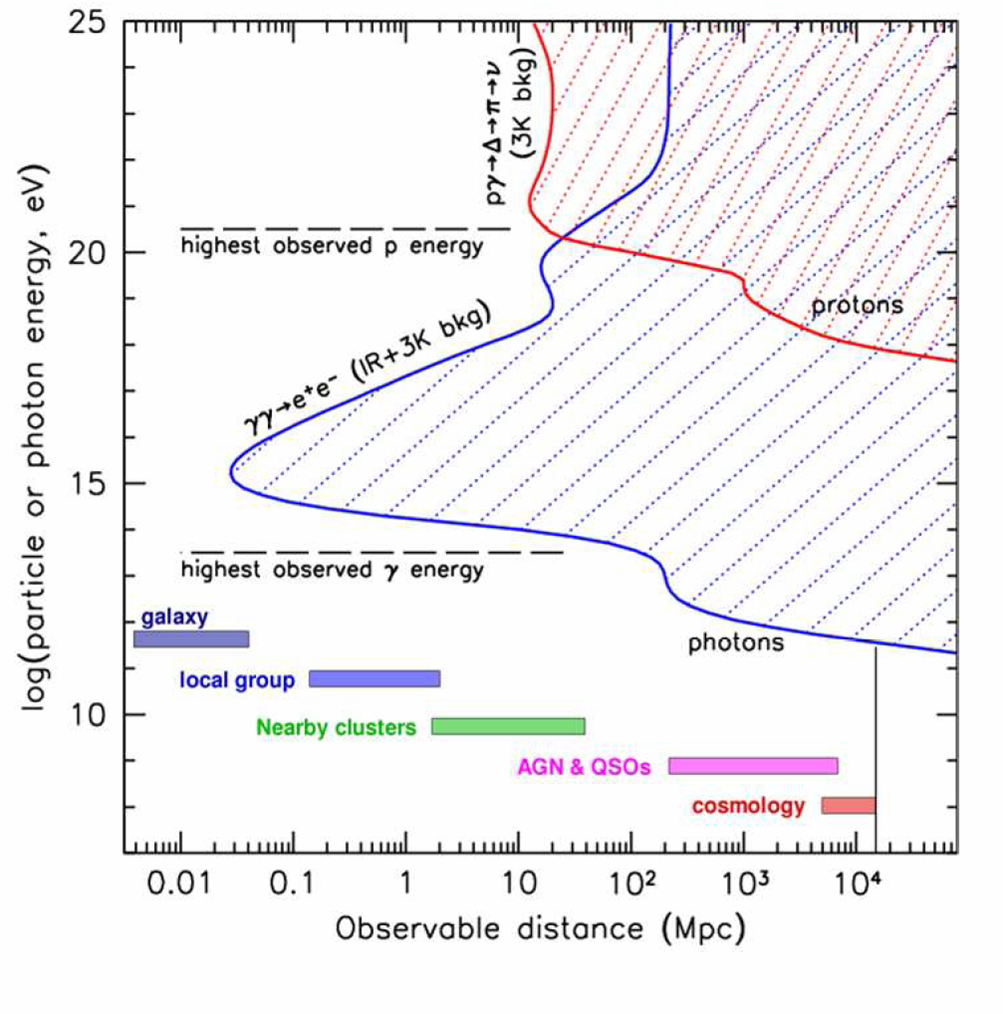}
   \includegraphics[width=3.5in]{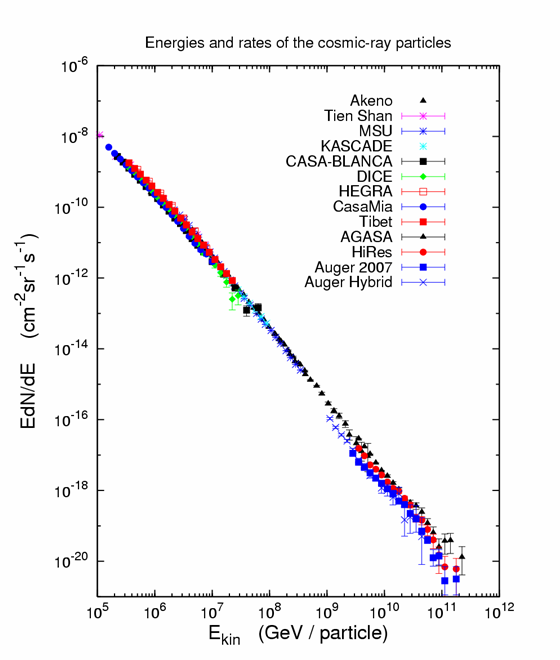}
    \caption{ Left:  Messenger particles energy vs distance they propagate. Photons (blue) are attenuated by interaction with the CMB and IR background radiation. Protons (red) interact with the CMB while undergoing magnetic deflection at lower energies~\cite{nu_prop}.  Right: The spectrum of cosmic rays.  The flux is shown as a function of energy~\cite{cr_spec}. }
   \label{fig:messengers}
\end{figure}

What information we do have on the astrophysical sources themselves has come from astronomical observations using photons which, because they are neutral, traverse space without bending and point back to their source. These measurements, which extend from the radio and infrared through high energy gamma-rays, have provided invaluable information in our attempt to assemble a comprehensive theory on the origin of cosmic rays and the acceleration mechanisms at work in high energy astrophysical objects. However, using gamma ray measurements alone, it is difficult to determine if the gamma rays are produced only by relativistic electrons or from hadronic processes that could also be the source of the galactic cosmic rays, since high energy electrons will produce synchrotron radiation and can transfer their energy to ambient photon fields through inverse Compton scattering.

The observation of neutrinos from a source would be unambiguous evidence for hadronic acceleration, since cosmic proton accelerators are expected to produce cosmic rays, gamma rays and neutrinos with similar luminosities.  This prediction arises from the conditions required for acceleration; a massive bulk flow of relativistic charged particles and a magnetic field to confine the charged particles within the acceleration region.  Candidate acceleration sites might be powered by large gravitational forces such as those found near black holes. 
Electrons accelerated in these regions would loose energy through synchrotron radiation thus, if protons were also being accelerated at the site, they would collide with the resulting radiation field,  producing mesons in processes such as
\begin{eqnarray*}
p+ \gamma &\rightarrow \Delta^+ \rightarrow \pi^0 + p \\
&\rightarrow \Delta^+ \rightarrow \pi^+ + n.\\
\end{eqnarray*}  
If neutral pions are produced, they will decay to high energy gamma rays whose energy will be shifted downward through pair production as they escape the source. Charged pions will decay mainly to a muon and a muon neutrino.  Although the protons may remain confined in the magnetic field, the neutrons and neutrinos  may escape, with the neutron subsequently decaying into a proton, an electron and a neutrino.  It is also possible to imagine a ``hidden source'' which acts as a beam dump from which only neutrinos could escape, however, such sources would not contribute to the observed cosmic ray flux.

 In recent years experiments such as Whipple, Milagro~\cite{Milagro} and HESS~\cite{HESS} have discovered a number of sources of very high energy gamma rays, both galactic sources and distant sources such as Active Galactic Nuclei.  Many of the gamma-ray emitters in our galaxy are associated with pulsar wind nebula and supernova remnants, which are generally believed to be the source of galactic cosmic rays. 
Estimates for neutrino fluxes for these galactic sources indicate they may be detectable by cubic-kilometer scale detectors~\cite{beacom, halzen2}. 

For astrophysical sources outside our galaxy the pervasive cosmic microwave background (CMB) and infrared (IR) radiation limits the propagation of gamma-rays and protons as messenger particles (Figure~\ref{fig:messengers} left).  Photons of a high enough energy will pair produce on the CMB and IR radiation limiting the horizon for TeV gamma rays to tens of Megaparsec.  Although protons can penetrate larger distances, because they carry an electric charge, the trajectories of protons are deflected by the magnetic field of our galaxy and do not point back to their sources.  However, at the high end of the proton energy spectrum  (above $\approx 10^{18}$ eV) this deflection becomes small enough that proton astronomy becomes possible. Currently operating extensive air shower arrays such as HiRes and Auger have exploited this fact in looking for the sources of ultra high energy cosmic rays.  Unfortunately, at these energies proton interaction with the CMB (the GZK effect~\cite{GZK1,GZK2}) again results in pion production and limits the propagation distance to tens of Megaparsec.

A steep sharpening of the cosmic ray energy spectrum has recently been observed by both the HiRes~\cite{hires} and Auger~\cite{auger} observatories at around $5 \times 10^{19}$ eV, which corresponds to the expected energy of the GZK cutoff (red line in Fig~\ref{fig:messengers}).  This could indicate that some of  the sources of the high energy cosmic rays have reached their maximum acceleration energy, or it could indicate that the highest energy cosmic rays are indeed  extragalactic in origin and the energy of the observed proton spectrum is being shifted downward as they propagate over long distances.  This raises several confounding questions.  Could the spectrum of energies produced at the source extend even higher?  Could the high energies cosmic rays we see be the product of decaying pions produced on the CMB? Since neutrinos interact only weakly, they can propagate indefinitely and still point back to their source.  In addition, as for the galactic sources, the presence of neutrinos is the hallmark of hadronic acceleration, providing insight into what fuels these most powerful astrophysical engines.  
Whatever the origin of cosmic rays, the acceleration of charged particles to the observed energies requires enormous gravitational forces such as those found in accreting black holes at the center of active galactic nuclei (AGN) or at the core of supernova remnants (SNR). 

Another candidate for cosmic ray acceleration sites are gamma-ray bursts (GRB).  Conventional models describe a rapidly expanding fireball of unknown origins where gamma rays in the fireball collide with high energy protons, creating a neutrino afterglow through similar mechanisms to those described above.  In fact, it is a tantalizing coincidence that the energy injection rate of photons integrated over all gamma-ray bursts is very similar to the energy injection rate of ultra high energy cosmic rays, (UHECRs) leading many to speculate that gamma-ray bursts and the source of UHECRs may be one and the same~\cite{waxman, halzen1}.

\begin{figure}[htbp] 
\begin{center}
\includegraphics[width=3.7in]{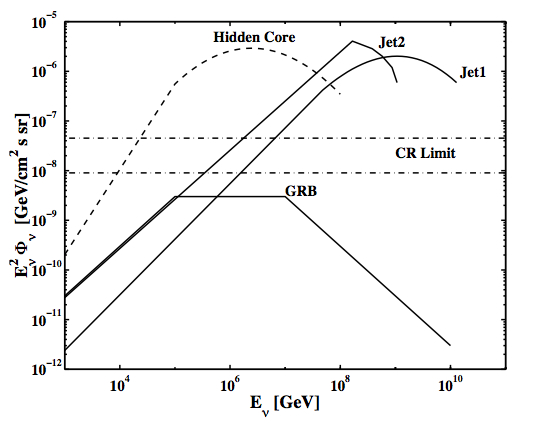}
\end{center}
 \caption{ The Waxman-Bahcall upper limit of the cosmological neutrino flux (dot dashed lines) based on the observed cosmic ray spectrum.  The solid lines labeled ``Jet'' show the predicted flux based on gamma-ray emission from blazars, while the GRB line is based upon models of GRB emission.  The dashed lines show predictions for neutrinos at the core of AGNs from which the parent protons cannot escape.~\cite{wb}}
 \label{fig:bound}
\end{figure}

If the same acceleration mechanisms are responsible for the observed spectrum of cosmic rays and gamma rays, the connection between the expected luminosities and neutrinos can be used to predict the neutrino flux, and ultimately, constrain proposed models of high energy phenomena such as AGNs and GRBs.  John Bahcall and collaborator Eli Waxman made a series of calculations connecting the expected diffuse flux of neutrinos to the observed high energy cosmic ray spectrum\cite{wb, wb2}. They claimed that the cosmic ray spectrum placed the most stringent limits on the flux of cosmic neutrinos from ``transparent sources", those from which neutrons may escape.   Note that the cosmic ray spectrum does not limit the flux from hidden sources.  Waxman and Bahcall  assumed a proton spectrum of   $dN_p/dE_p \propto E^{-2}$ at the source, since this spectrum is typical of Fermi shock acceleration.  They then calculated the energy production rate in protons at the source assuming that the same sources produced the observed spectrum of cosmic rays.  From particle physics, they were then able to deduce the expected energy density of muon neutrinos.  Their upper bound is shown by the dash-dotted line in Figure~\ref{fig:bound}.  This limit contradicted several AGN models labeled ``Jet" in the same figure.
Noting that most of the extragalactic gamma ray energy is contained in a diffuse glow rather rather than from discrete sources suggests another approach; assuming a particular spectrum at the source, and normalizing it to the observed diffuse flux of gamma rays.
More recently, a  calculation of the expected neutrino flux based on the cosmic ray spectrum has been made by Mannheim, Protheroe, and Rachen~\cite{mpr} which did not assume a particular injection spectrum but instead assumed a specific spectrum for the observable cosmic ray flux, taking into account  production, source evolution, and propogation effects, that yielded a more stringent limit than those of Waxman and Bahcall.  In any case, a neutrino observatory with an effective area of 1 ${\rm km^3}$ will be sensitive to diffuse glows of neutrinos that are well below the upper bounds calculated by any of these methods.

In summary, the motivation for pursuing high energy neutrino astrophysics is the unique and complementary role that neutrinos could play in our observation of the high energy universe. The observation of neutrino emission from astrophysical sources would provide unambiguous evidence for hadronic acceleration and the source of cosmic rays. The neutrino flux estimates discussed above indicate that neutrino detectors of cubic kilometer scale are necessary to approach the sensitivity needed. Furthermore, neutrinos propagate through the universe almost undisturbed to the highest energies without deflecting in  magnetic fields and are therefore usable as astronomical messenger particles at energies where gamma-rays or protons are not, and therefore hold the promise of opening a new energy window on the high energy universe.

\section{Neutrino Telescopes}

\subsection{Water/Ice Cherenkov technique}


It is the fact that neutrinos interact only weakly that gives them the unique ability to probe some of the most enigmatic processes at the core of stars and galaxies from across cosmological distances, but this is a mixed blessing, since it makes neutrinos very difficult to detect, even at high energies where the cross sections are more favorable.  Neutrinos can only be sensed indirectly after they interact with a target nucleus.  In the inverse beta decay process, the incoming neutrino is captured on a target nucleus when a $W$ boson is exchanged (a ``charged current'' interaction).  The target atom undergoes a nuclear transmutation, while a charged lepton exits the interaction point in place of the incoming neutrino.  The tedious method of counting the number of transmuted nuclei that was employed by Ray Davis is not well suited for high energy neutrino telescopes since it yields no directional information, and it limits the target  to the volume enclosed by the tank.  

\begin{figure}[htbp]
\begin{center}
\includegraphics[width=3.0in]{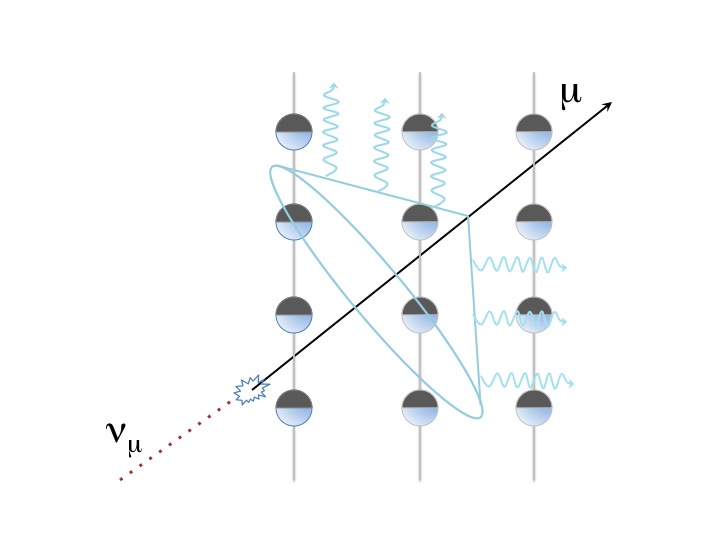}
\hspace{-.5in}
\includegraphics[width=3.0in]{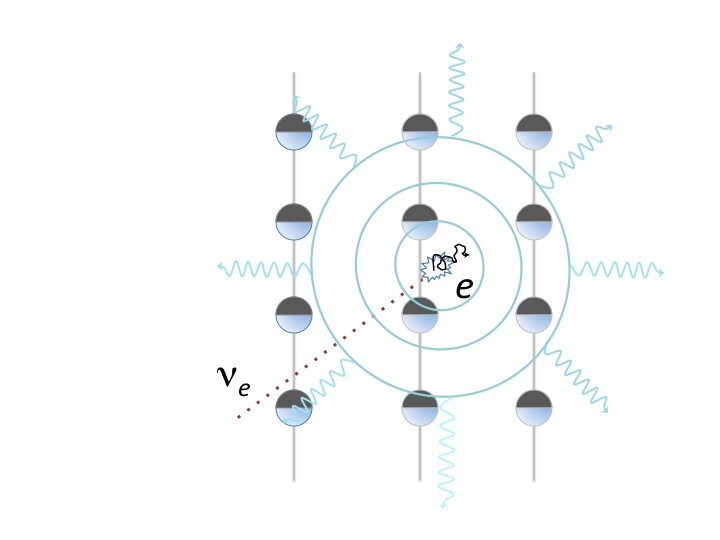}
\caption{Schematic of a  neutrino interaction and the subsequent detection of the secondary lepton in a water/ice Cherenkov detector.  The expanding surface of constant phase is shown.  The Cherenkov light is detected and time stamped by optical modules embedded within the volume of the Cherenkov radiator, so that the track or cascade may be reconstructed.}
\label{fig:cherenkov}
\end{center}
\end{figure}

More recently, instruments such as  Super-Kamiokande  have focussed on the detection of the secondary leptons, which emit Cherenkov radiation when traversing optically clear media at relativistic speeds and can therefore be reconstructed using embedded photomultiplier tubes.  At high enough energies, the secondary lepton will emerge approximately colinear with the incoming neutrino allowing the origin of the parent neutrino to be mapped.  This is particularly advantageous in the detection of muon neutrinos since secondary muons produced in a high energy neutrino interaction can propagate for a kilometer or more, allowing the detection of tracks originating well outside of the instrumented volume.  The event topology for a muon neutrino interaction is shown in Figure~\ref{fig:cherenkov} left.    Although Cherenkov telescopes have the greatest sensitivity to muon neutrinos, they are also sensitive to other flavors.  The showering induced by electron neutrinos limits their detection to interactions near or enclosed within the instrumented volume (while a muon's path must only intersect the instrumented volume, thus the sensitivity is related to the instrumented area).  Electron showers also yield poor directional information.  Figure~\ref{fig:cherenkov} right shows the evolution of the cascade initiated by an electron neutrino interaction. A distinctive ``double bang'' signature results from tau neutrino interaction in cases where the tau has sufficient energy to travel a large enough distance from the point of production to produce two distinct cascades, one at the point of production and one at the point of decay. The water Cherenkov technique, which had been previously employed in the search for proton decay, has become the mainstay of modern neutrino astronomy.

By the time Supernova 1987A was observed, efforts to develop instrumentation for an astrophysical neutrino observatory with a much higher energy threshold had already been underway for more than 10 years.  
Unlike detectors used in the study of solar or atmospheric neutrinos which were successfully carried out using manmade tanks, the sheer size of the instrumented volume needed  limits these telescopes to the approach, pioneered by DUMAND, Baikal and AMANDA, of using large natural reservoirs of water or clear ice which can serve as both a target volume and a Cherenkov radiator. Contemporary efforts to build a neutrino observatory owe a great debt to the 20 years of groundwork laid by these pioneering collaborations.

The Deep Underwater Muon and Neutrino Detection Project (DUMAND) was the first to construct an operational prototype when, in 1987, it deployed a string of photomultiplier tubes tethered to a ship in the ocean near the big island of Hawaii and made observations using the ocean water itself as a target medium. The success of this early test led to a proposal for a permanent detector array consisting of nine strings moored on the ocean floor at a depth of 4800m.  Unfortunately, only one of these strings was ever deployed, and a leaky connector forced the team to retrieve the components for repair after only 10 hours of observation~\cite{Wilkes94}. Funding to DUMAND was cut before a design was fully realized, and the refurbished string was never redeployed. An excellent review of the history of DUMAND and early neutrino telescopes is available in~\cite{roberts}.

The Baikal group overcame many of the costs and risks associated with deep-sea deployment by choosing to construct their observatory in a Siberian lake rather than open ocean.  New strings were added during the winter season, when the thick ice layer that formed on the surface of the lake provided a stable platform for installation. Like DUMAND, the group had to overcome early problems with leaky connectors, but these problems were solved with a custom connector designed to survive pressures down to 1.6 km of water, and the first 36 optical modules (NT36) were installed in 1993. By 1998 the Baikal detector NT200 was completed with 192 optical modules and it continues to operate. The Baikal collaboration had successfully demonstrated the Cherenkov technique for the detection and reconstruction of high energy neutrinos with a large sparse detector array in deep open water~\cite{baikal}.

The Cherenkov technique in naturally occurring deep clear ice was first demonstrated around 1997 by the Antarctic Muon and Neutrino Detector Array (AMANDA) collaboration, which deployed 302 optical modules in the deep ice sheet covering the South Pole. With the construction of this array, the AMANDA collaboration established the technique of using a high pressure, hot water drill for installation, and carried out detailed surveys of the optical properties of the Antarctic ice. By the year 2000 the AMANDA collaboration had installed 677 optical modules beneath the South Pole at depths of 1500-2000m in an array of diameter 200m. This detector continues to operate and, to date has produced the most sensitive searches for astrophysical neutrinos in the TeV range.

The Cherenkov technique for detecting high energy neutrinos has now been well established in both water and ice by the Baikal and AMANDA detectors respectively. 
The steeply falling cosmic ray spectrum, which gives an estimate of the expected neutrino flux as outlined in Section 1.1, sets a scale of a cubic kilometer for high energy neutrino observatories to reach a sensitivity of astrophysical interest.   
These successes have paved the way for construction of the current generation of discovery scale instruments, which include ANTARES, a detector of AMANDA scale in the Mediterranean Sea, with planning for an eventual cubic kilometer detector, and IceCube, a detector of cubic kilometer scale at the South Pole. 
 Table~\ref{tab:telescopes} summarizes the basic characteristics of the world's neutrino telescopes.

\begin{table}[t]
\caption{\label{tab:telescopes}Optical Cherenkov neutrino telescopes currently operating or under construction.}
\begin{indented}
\item[]\begin{tabular}{llccc}
\br
&&&instrumented&date of\\
&location&phototubes&area (${\rm km}^2$) & operation\\
\br
{\bf Lake Baikal}&Siberia & & & \\
NT36,72,96 &      & 36,72,96 & & 1993 \\
NT200& & 192 &  .002 &  1998 \\
\mr
{\bf AMANDA} & South Pole & & & \\
AMANDA B-10 &          & 302 & .01 & 1997 \\
AMANDA II & &  677 &  .03 &  2000 \\
\mr
{\bf IceCube} &South Pole    && & \\
IC-9&            &540& .1  &2006 \\
IC-22 & & 1320 & .25  & 2007  \\
IC-40 & & 2400 &  .5 & 2008\\
IC-80 & & 4800 & 1  & 2011\\
\mr
{\bf ANTARES} & Mediterranean& 900 & .03 & 2008 \\
{\bf Nestor} & Mediterranean& & & R\&D \\
{\bf Nemo} & Mediterranean & & & R\&D \\
{\bf KM3Net} & Mediterranean & & 1 & Design Phase \\
\br
\end{tabular}
\end{indented}
\end{table}

\subsubsection{Performance requirements}

\begin{figure}[htbp]
\begin{center}
\includegraphics[width=4.0in]{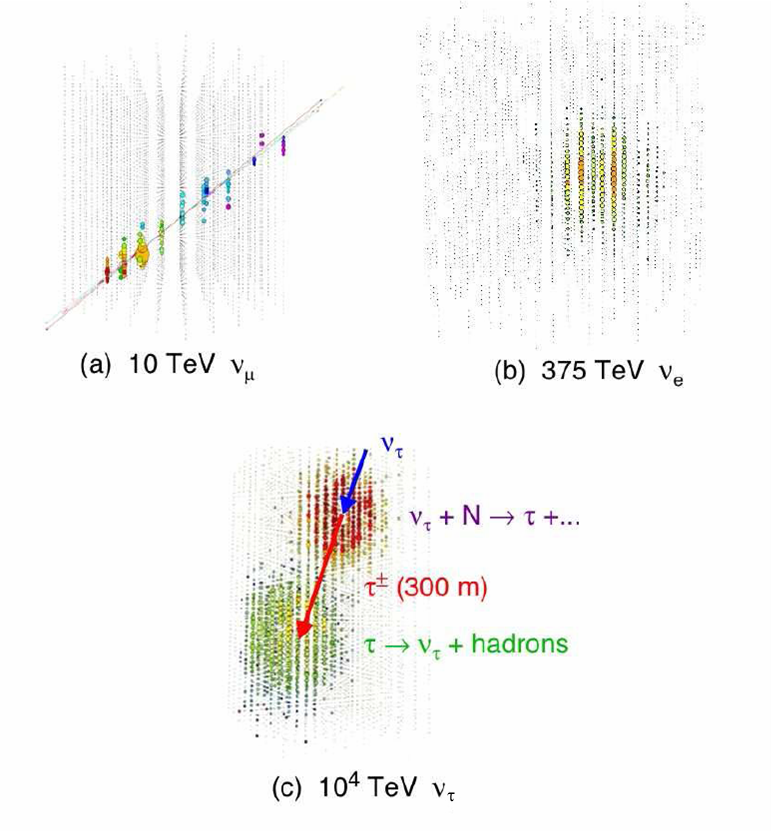}
\caption{Simulation of event signatures for three flavors of neutrino interaction in the IceCube detector. (a) Shows the muon track fro a 10 TeV muon neutrino interacting out side the detector volume. (b) Shows a 375 TeV electron neutrino interacting in the detector volume. The secondary electron immediately showers and the resulting light spreads outward almost uniformly at large distances due to the scattering of the light. (c) A 10 PeV tau neutrino interaction showing the "double-bang" signature.}
\label{fig:i3particles}
\end{center}
\end{figure}

The functional unit of any water Cherenkov telescope is the photomultiplier tube (PMT).
To ensure high signal efficiency, the tubes must be sensitive to very low light levels and have low dark noise rates.
  In $\nu_\mu$  charged current interactions, the direction of the secondary muon is correlated with the direction of the parent to within 1 degree or better for neutrinos with an energy above 1 TeV. 
  In order to reconstruct direction of the secondary muon to this accuracy, the PMTs must measure the relative arrival time of the Cherenkov light to within nanoseconds.  In addition, the brightness of the event is correlated to the energy of the muon, since at these high energies, muons are not minimum ionizing but lose energy through radiative processes, therefore the energy of the event can be estimated by measuring the amount of light in the detector.
  The $\nu_e$ and $\nu_\tau$ interactions produce secondary electrons and taus, which appear as shower-like events. Showers have poor directional resolution but provide more accurate energy resolution if they are contained within the detector volume.  Figure~\ref{fig:i3particles} shows a simulation of how these three topologies would appear in the completed IceCube detector.  Although the Cherenkov technique is well established now, the performance requirements needed to accomplish the challenging science goals, especially given the uncontrolled natural environments in which the observatories are located, and the sheer scale of the instruments themselves, make the design and construction challenging.

Neutrino telescopes are triggered almost entirely by the prolific flux of downward going muons generated by primary cosmic rays in the atmosphere. Although the atmospheric background decreases considerably with depth,  the cleanest muon neutrino signatures are obtained by searching for muons traveling upward through the earth where the cosmic ray muons are completely filtered by the earth.  In order to reduce the background of the downgoing muons mis-reconstructed as upward going to an acceptable rate, a rejection rate of $10^6$ is required, and has clearly been demonstrated by existing detectors. The downgoing  muons produced in cosmic ray interactions can be used as a testbeam for calibration of the instrument. 

Muon neutrinos produced in the atmosphere  give rise to an irreducible, isotropic neutrino background in any neutrino observatory, since atmospheric muon neutrinos readily penetrate the earth.   While neutrino observatories have the greatest effective area and superior pointing resolution for the muon channel,  the signatures of tau neutrinos and electron neutrinos are nearly background free and may be sought over the entire solid angle. 

Because the technique of selecting upgoing muon events restricts the view of muon neutrinos to half the sky, telescopes at the South Pole and the Mediterranean are complementary.  From their vantage point at the South Pole, the IceCube and AMANDA arrays provide a fixed view of the Northern celestial hemisphere, while detectors in the Mediterranean such as ANTARES continually sweep out the Southern sky, with the galactic center in view 63\% of the time. The differing properties of cold ice and sea water also give each instrument unique strengths.  Deep sea water does not have the dust and bubbles found in the ice, which serve as scattering sites, and therefore provides superior angular resolution. Ice has a longer absorption length allowing sparser instrument spacing and therefore a higher energy threshold for a fixed number of PMTs. The ice also has no backgrounds from bioluminescence and radioactivity and provides for very low thermal noise. 

\subsubsection{South Pole telescopes}

\begin{figure}[htbp]
\begin{center}
\includegraphics[width=3.5in]{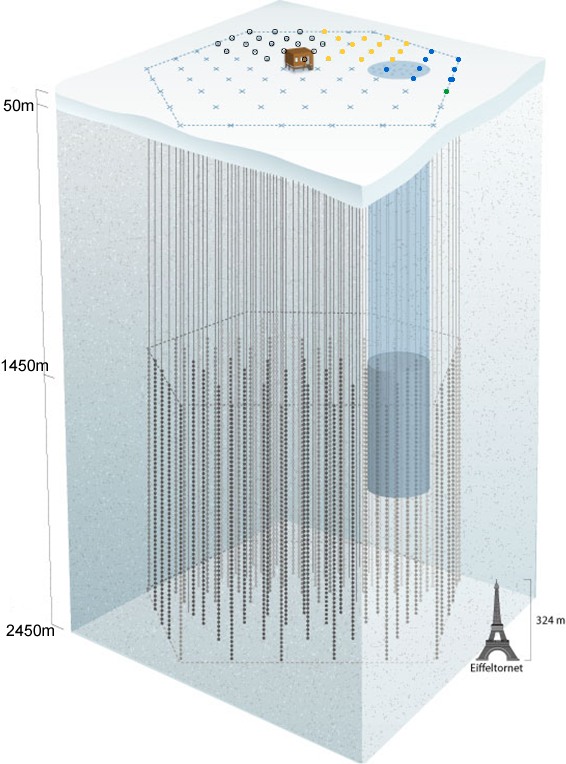}
\caption{ A schematic of the IceCube detector array. IceCube will have 80 strings with 60 DOMs on each string at a depth of 1450m to 2450m instrumenting nearly 1 cubic kilometer of ice under the Antarctic South Pole. Each of the 80 strings also has 2 surface tanks instrumented with 2 DOMs each to form the IceTop air-shower array above.  All of the tanks and strings are controlled and read out from computers located in the IceCube laboratory (center).  The locations of strings that were already installed and operational by early 2008 are shown by the colored dots, with the colors showing the year of installation (green for 04-05, blue for 05-06, yellow for 06-07, and open dots for 07-08). 
}
\label{fig:i3schematic}
\end{center}
\end{figure}

Following  the successful operation of the AMANDA detector, construction of a cubic kilometer array, IceCube, commenced in the 2004-2005 austral summer. When completed, IceCube will consist of an in-ice detector with a baseline design of 80 strings of digital optical modules (DOMs) in a 125 m triangular lattice (Figure~\ref{fig:i3schematic}).  Each string will supply power, communications, and mechanical support to the 60 DOMs attached at 17 m intervals along its length between 1450 and 2450 m depth for a total 4800 DOMs in the ice.  Each DOM consists of a photomultiplier tube and electronics for power distribution, timing calibrations, and data acquisition enclosed within a glass pressure sphere.   Two surface tanks are located near the top of each string for background veto, calibration studies, and air shower studies and comprise the IceTop surface detector. 
Finally, the old AMANDA array is embedded within the IceCube array and is integrated into the IceCube readout and reconstruction. Its smaller string spacing provides  for improved sensitivity at lower energies. 

Recently, a proposal to install an additional 6 strings near the center of IceCube was approved. The ``deep core'' will more densely instrument a volume within IceCube centered in radius and at the bottom of the array  (depth of 2000m to 2400m) where the ice is exceptionally clear and dust free.
The smaller sensor spacing, larger overburden and the use of IceCube DOMs as a veto will allow for the reconstruction of short tracks that start and stop in the detector (contained events) that are induced by  low energy neutrino interactions with minimal background from atmospheric muons.  In addition to extending the low energy sensitivity of IceCube, the deep core would enable the use of a contained event topology to extend the sky coverage at low energies to the southern sky, including the galactic center. 

\begin{figure}[htbp]
\begin{center}
\includegraphics[width=4.0in]{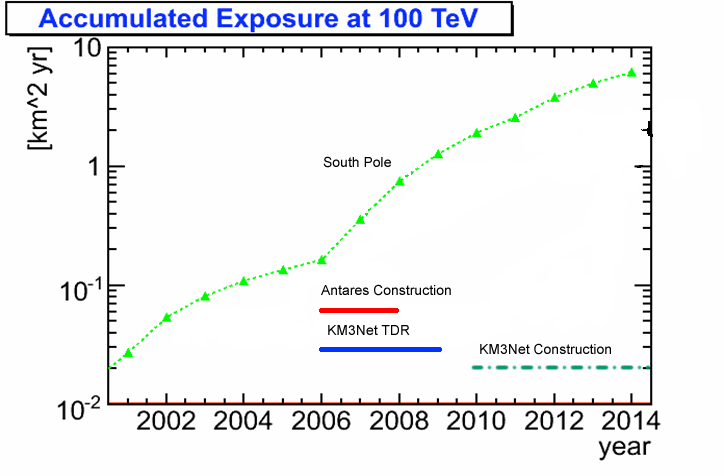}
\caption{The accumulated exposure in units of instrumented kilometer-squared years versus calendar year for the AMANDA and now AMANDA plus IceCube South Pole detector arrays. Also shown is the time scale for ANTARES construction and an estimate for the KM3NeT design phase in the Mediterranean.}
\label{fig:exposure}
\end{center}
\end{figure}

IceCube deploys incrementally during the austral summer seasons (November-February)
and operates the array over the intervening period until the next deployment season.  Table~\ref{tab:telescopes} shows the configurations in which IceCube has been operated since it began pre-operations with 9 strings in 2006.
With the 18 strings added in the most recent austral summer (07-08), the total number of deployed strings presently stands at 40 and the telescope is now half complete.  
The performance of the detector has thus far exceeded expectations with less than a 1\% failure rate for the installed optical modules, and smooth and reliable detector operations.  Preliminary studies show that a median angular resolution of 0.5 degrees for muon tracks from an $E^{-2}$ spectrum can be achieved for tracks with  horizontal projections of 1 km in the detector (slightly worse for vertical events).  The data being taken is already of sufficient quality for analysis, and the first physics results from the 9 string configuration are  already available.  (See Section 3.)
Deployments will continue in each austral summer until the array is completed in the year 2011.  
The accumulated exposure for the South Pole neutrino telescopes is shown in Figure~\ref{fig:exposure} and is shown to be rapidly increasing, reaching 1 ${\rm km}^2 \cdot {\rm yr}$ in 2009.

\subsubsection{Mediterranean telescopes}

 The ongoing neutrino telescope initiatives in the Mediterranean are shown in Table~\ref{tab:telescopes}, including ANTARES~\cite{antares}, NESTOR~\cite{nestor} and NEMO~\cite{nemo}. In addition, the ${\rm km^3}$ NEutrino Telescope, or KM3NET~\cite{km3net}, is a planned cubic-kilometer scale detector in the Mediterranean, which will build on the experience from the current three initiatives there. ANTARES has begun initial data collecting with a partial array and is near completion.

\begin{figure}[htbp]
\begin{center}
\includegraphics[width=4.5in]{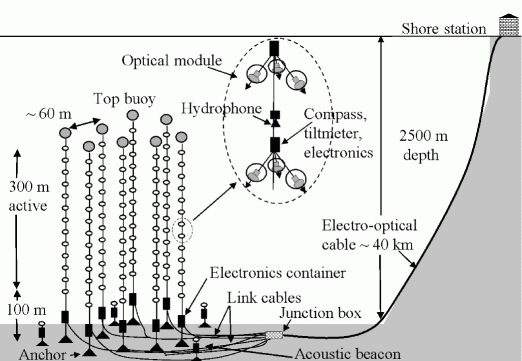}
\vspace{.2in}
\caption{Schematic of the Antares detector array.}
\label{fig:antaresschematic}
\end{center}
\end{figure}

The ANTARES (Astronomy with a Neutrino Telescope and Abyss environmental RESearch) telescope under construction in a deep Mediterranean location off the coast of France is shown schematically in Figure~\ref{fig:antaresschematic}.  It was completed on May 30, 2008 and  has 12 strings, each with 75 optical sensors, configured as 25 triplets, for a total of 900 optical modules at a depth of 2500m. 
Ten complete strings have been operating since December 2007, and  five have been connected and operational since Jan 2007. 

\begin{figure}[htbp]
\begin{center}
\includegraphics[width=2.75in]{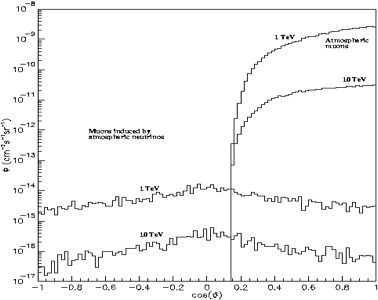}
\includegraphics[width=3in]{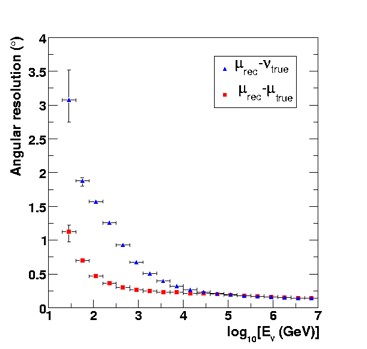}

\caption{Simulated angular flux of events in the ANTARES detector. Atmospheric muons are only seen above the horizon, whereas muons induced by neutrinos produced in atmospheric interactions are distributed nearly isotropically. The flux is seen to drop precipitously with energy  for both distributions.~\cite{antares1} The angle between the true reconstructed muon directions (squares) and the true neutrino direction and reconstructed muon directions (triangles) in ANTARES.  At energies below 1 TeV the true separation between the direction of the muon and the parent neutrino dominates the uncertainty in the neutrino direction.~\cite{antares1}}
\label{fig:updown}
\end{center}
\end{figure}

\begin{figure}[htbp]
\begin{center}
\includegraphics[width=2.0in]{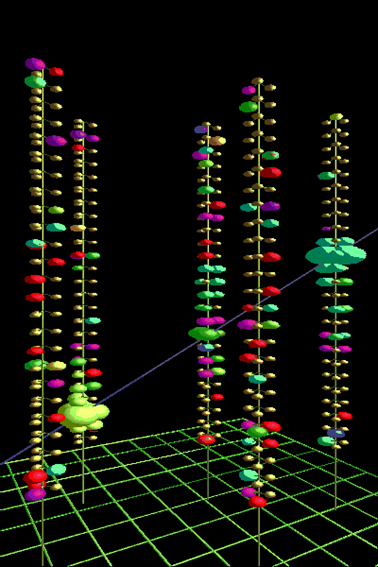}
\hspace{.3in}
\includegraphics[width=3.5in]{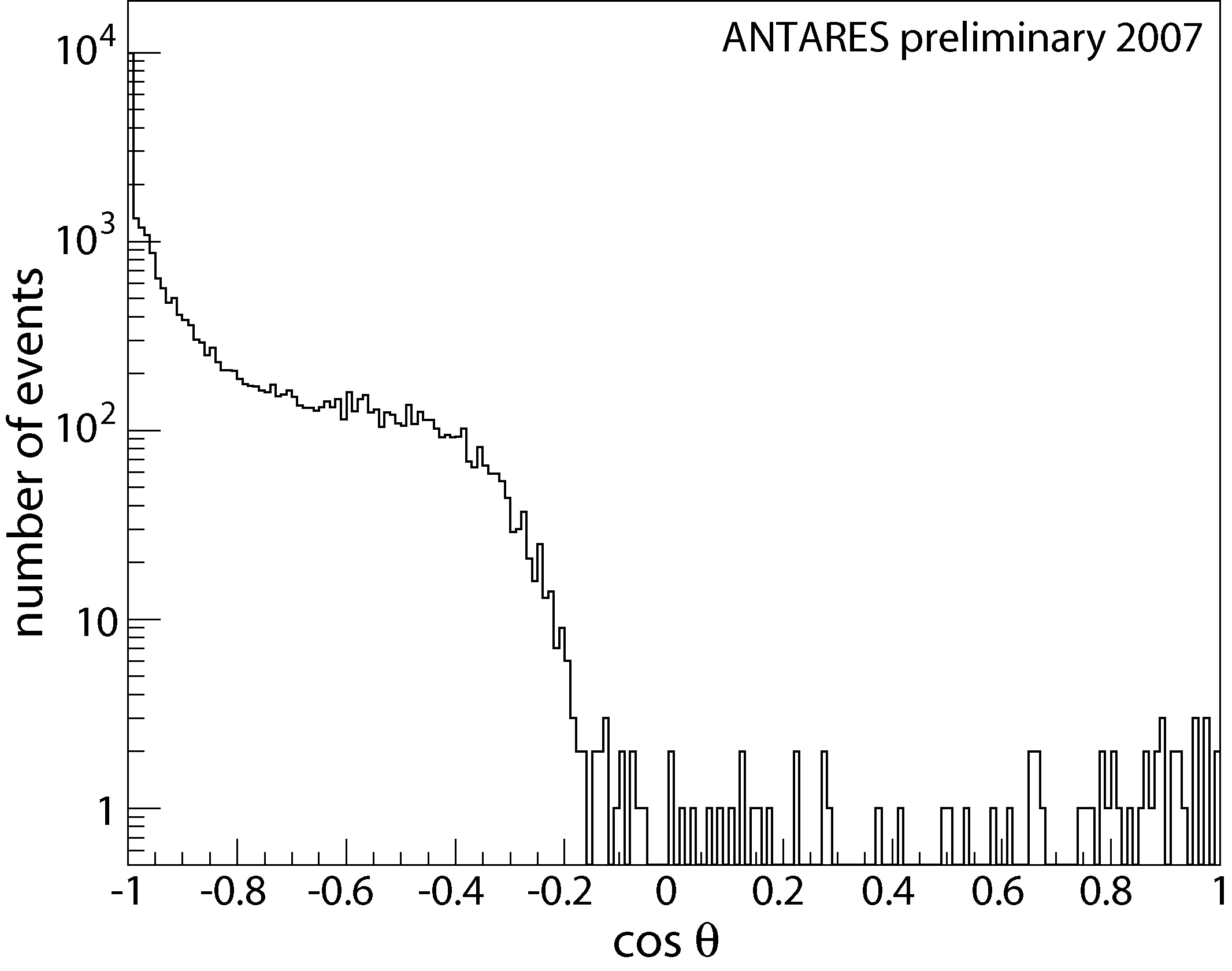}
\caption{Left: a neutrino candidate event from the ANTARES neutrino telescope in the Mediterranean sea. Right: The angular distribution of tracks reconstructed in the partially completed ANTARES array. The distribution shows the large rate of downward going muon tracks and the clear upward going neutrino signal. Neutrino events near the horizon are also discernible, showing the power of the excellent angular resolution achieved.~\cite{antares2}}
\label{fig:antaresevent}
\end{center}
\end{figure}

As described above, the relative absence of dust and bubbles in water detectors provide for less scattering of the Cherenkov light, which in turn gives rise to the detection of more direct photons (i.e. photons not scattered before detection) and an excellent angular resolution. Studies of the performance of the first five deployed lines show a timing resolution of  $<0.4$ ns~\cite{antares2}.  The finished array will have an angular resolution of $<0.3^o$~\cite{ant_ps} at high energies, while at low energies, the uncertainty in direction is dominated by the angle between the parent neutrino and the muon track.  Figure~\ref{fig:updown} shows results from a simulation for the ANTARES detector. On the left is the distribution of muon tracks predicted for the ANTARES detector. Note the extremely rapid fall off in the reconstructed direction for the downward going atmospheric muon sample near the horizon. This is due to the excellent angular resolution of the ANTARES detector (Figure~\ref{fig:updown}, right). Shown in the left of Figure~\ref{fig:antaresevent} is an upward going neutrino event candidate in the ANTARES detector. The plot in right of the figure shows the angular distribution of reconstructed tracks in their array. The distribution shows a clear upward going neutrino signal. ANTARES has recently eclipsed Baikal as the most sensitive neutrino detector in the northern hemisphere.

\subsection{Techniques for higher energies}

The recent observation of a steepening of the cosmic ray spectrum at ultra high energies by both the High Resolution Fly's Eye (HiRes)~\cite{hires}  and Auger~\cite{auger} observatories may be evidence for the GZK cutoff.
As discussed in Section 1.1, this could be a strong indication that the highest energy cosmic rays are indeed extragalactic in origin.
This also implies a guaranteed flux of neutrinos which could be used to test models of cosmic ray production at ultra high energies.
However, due to the low flux of cosmic rays with energies
at the GZK scale, kilometer scale neutrino arrays are not expected to net more than one GZK neutrino
per year despite their formidable size.  Their sensitivity is limited chiefly to the TeV to PeV energy range.  It  can be calculated from the fluxes in the observed cosmic ray spectrum that one would  need an array covering on the order of 100 square kilometers to record a statistically significant
sample of GZK neutrinos for study.  The attenuation length of light in ice and water,
which is on the order of 200m, makes the conventional technology too costly to scale up to this size.
Existing extensive air shower arrays such as Auger and HiRes have some sensitivity to showers produced by earth-skimming neutrinos in the EeV range.
In addition, two new approaches are being pursued to achieve significant sensitivities at energies of $10^{16}$ to $10^{20}$ eV:  detecting RF emission or acoustic waves, both of which would result from an ultra high energy particle cascade.

\subsubsection{Extensive air shower arrays}

At extremely high energies, the Earth becomes opaque to neutrinos, thus neutrino induced cascades would appear  at or just below the horizon.  At these energies, a secondary tau produced in a neutrino interaction can travel tens of kilometers in the Earth emerging from the ground nearly parallel to the horizon.  Although tau neutrinos are not expected to be produced in astrophysical sources or from the interactions of cosmic rays on the CMB, a ratio of 1:1:1 for $\nu_e:\nu_\mu:\nu_\tau$ is expected at Earth from cosmogenic sources due to neutrino oscillations over long baselines. Similarly, the Landau-Pomeranchuk-Migdal (LPM) effect~\cite{LPM} will decrease the cross section for bremsstrahlung and pair production of high energy charged particles in dense media, causing electron showers to evolve more gradually, allowing a measurable number of charged particles  to emerge from the Earth even for a cascade that was induced by a neutrino interaction several kilometers below the surface. 
Extensive air shower arrays can distinguish a shower induced by neutrino secondaries coming in from at or below the horizon from one initiated by a vertical cosmic ray shower from the direction and the age of the air shower.

\subsubsection{Askaryan radio telescopes}

The idea that high energy charged particle showers might produce coherent radio emission in dense media was first suggested in a 1962 paper by Soviet physicist G. Askaryan~\cite{askaryan, askaryan2} where he proposed that these emissions could be used to monitor the moon for cosmic rays impinging on its surface.  These emissions would arise  
as an excess of negative charge builds up as electrons are swept out along the relativistically advancing shower front.   The longer wavelength components of the broadband radiation from the motion of this  large net negative charge will add coherently, while in the smaller optical wavelengths, the electric field vectors add incoherently, giving rise to a radio frequency impulse.   This effect, which now bears his name, was demonstrated recently in fixed target experiments by directing electron pulses into beam dumps of sand, salt, and ice~\cite{ask_sand, ask_salt, ask_ice}.  
The shower dimensions determine the wavelengths over which these emissions are coherent, and in these dense media, the amount of power radiated in radio exceed the power radiated in the optical for shower energies greater than about $10^{15}$ eV.  These media also have attenuation lengths on the order of one kilometer for radio frequencies, thus the low
density of instrumentation required to monitor a large target volume for radio emissions may make a GZK scale telescope feasible.

\begin{figure}[htbp]
\begin{center}
\includegraphics[width=6in]{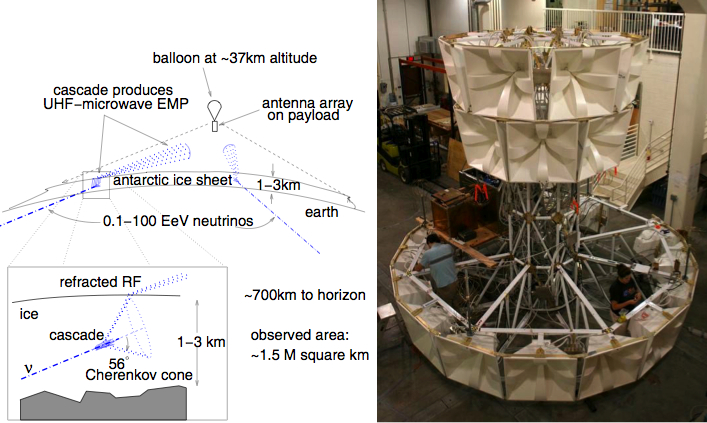}
\vspace{.2in}
\caption{Right: Schematic of the geometry of the ANITA event detection.  Radio emissions from Cherenkov showers below the critical angle refract out of the ice. On the left, the ANITA balloon payload, consisting of a horn antenna array, is shown. From~\cite{silvestri}.}
\label{fig:anita}
\end{center}
\end{figure}

Several current and proposed experiments exploit the Askaryan effect in the search for astrophysical neutrinos.  Perhaps the most ambitious to date is the ANtarctic Impulse Transient Array (ANITA)~\cite{anita}, a balloon borne antenna array which, for a few weeks in the austral summer of 06-07, surveyed the entire continent for radio frequency emissions emanating from horizontal neutrino induced showers that are refracted at the surface of the ice.   A schematic of the ANITA concept and a photograph of the apparatus is shown in Figure~\ref{fig:anita}. From its vantage point, ANITA had a clear line of sight over a vast 1.5 million square kilometers of the Antarctic ice cap,  however, the length of the observation time was limited to the duration of the balloon flight.  The results of ANITA's first flight will be available shortly. 
There is therefore a clear motivation for the development of a more permanent antenna array to monitor the ice for signals of GZK neutrinos.   One such array has been built by the Radio Ice Cherenkov Experiment (RICE) Collaboration~\cite{rice} and installed permanently within a 200m x 200m x 200m volume at depths from 100-300m near the AMANDA array at the South Pole.    Although it is unlikely that a small array such as RICE will ever detect a high energy neutrino, for the last 10 years it has occupied a unique niche, spanning the energy regime between AMANDA and ANITA. Much as DUMAND and Baikal paved the way for larger optical Cherenkov devices, these experiments have been invaluable in establishing the feasibility of a radio array, as well as the RF properties of South Pole ice. 

Several efforts are underway to develop the technology needed to build a permanent radio array
large enough to collect a statistically significant number of events at high energies.  
One proposed idea, put forth by the ARIANNA (Antarctic Ross Ice Shelf ANtenna Neutrino Array) Collaboration, is to use the Ross Ice Shelf as a waveguide for RF emissions from impinging neutrinos~\cite{arianna}.
IceCube is also studying several scenarios for extending the sensitivity of their array to higher energies,
in hopes of maturing the technology in time to propose an embedded GZK scale array shortly after IceCube has embarked on a program (AURA, the Askaryan Under ice Radio Array) to develop in-ice radio instrumentation for a large distributed array which could be installed below the surface of the ice. 
If such an array could be built at the South Pole surrounding the IceCube, the two arrays would form a hybrid observatory which would provide invaluable cross calibration information for a small subset of events.  To date, three RF test stations have been deployed~\cite{aura2} with another four planned for the austral summer of 08-09. Because of the high cost of drilling, a number of radio test stations are also being deployed near the surface (IceRay) to evaluate the compromise between in-ice and surface stations.

\subsubsection{Acoustic arrays}

Acoustic waves are generated from the thermal expansion when an ultra high energy lepton deposits massive amounts of energy in dense media such as ice or water.  The prospect of using the acoustic signal in the detection of high energy cascades is still in early development,  but early investigations by  SAUND~\cite{acoustic} (Study of Acoustic Ultra-high energy Neutrino Detection) to deploy hydrophones in ocean water,  and SPATS~\cite{spats} (South Pole Acoustic Test Setup) to install piezoelectric transducers in ice,
are encouraging.  Both ANTARES~\cite{ant_ac} and IceCube are actively pursuing acoustic sensors to extend their sensitivity to higher energies.

\section{Current Results and Future Prospects of the Existing Telescopes}

Here we present the the most important results to date from neutrino astronomy along with estimates of the improvement in sensitivity  expected from the new generation of detectors that are just beginning operations.

\subsection{Atmospheric neutrinos}

Neutrinos are copiously  produced as decay products of hadronic showers induced by the constant flux of cosmic rays hitting the atmosphere.  The distribution of these neutrinos has been measured repeatedly in neutrino telescopes starting with SuperKamiokande~\cite{sk-atm}.  Their isotropic production makes them invaluable as a testbeam for long baseline neutrino studies, since their path length through the Earth is zenith angle dependent.  In fact, a zenith angle dependent deficit of muons measured by SuperKamiokande provided some of the first evidence to corroborate the neutrino oscillation hypothesis, and it has since served as a testbed for more exotic neutrino oscillation scenarios, such as the violation of Lorentz Invariance which at high energy can distort the atmospheric spectrum~\cite{kelley}.  Despite their utility in oscillation studies, atmospheric neutrinos comprise an irreducible background in the search for astrophysical neutrinos.  For these reasons, the flux and energy distributions of atmospheric neutrinos continue to be studied with great interest, both to reduce the uncertainty in their spectrum, and for the potential to discover new physics as larger  telescopes provide the sensitivity  to probe subdominant effects.

\begin{figure}[htbp]
\begin{center}
\includegraphics[width=3.0in]{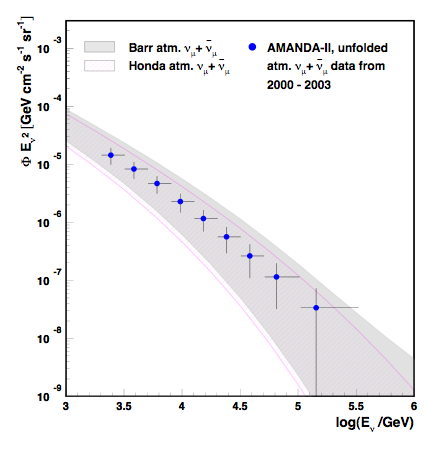}
\caption{The energy spectrum of atmospheric muon neutrinos measured in AMANDA II from 2000-2003~\cite{am_atmnu} compared to theoretical predictions (white and shaded bands).}
\label{fig:am_atmnu}
\end{center}
\end{figure}

Neutrino telescopes  do not provide a direct measure of  the energy of muon neutrino since the secondary muon may not deposit all of its energy within the instrumented volume.  However, with simulations, the energy of the parent spectrum of a sample of muons can be unfolded through the use of correlated variables.  The atmospheric muon  neutrino spectrum has been measured with AMANDA by using a neural network to compare a number of observables in data to those of simulated of monoenergetic neutrinos.   The resulting spectrum is shown in  Figure~\ref{fig:am_atmnu} where the bands show the theoretical flux expectations, which have a 30\% uncertainty.

IceCube's first physics result~\cite{ic_atmnu} was a measurement of the flux and distribution of atmospheric muon neutrinos based on the 137.4 days of livetime collected during 2006 with the 9 string configuration.  Because the atmospheric neutrino signal has already been extensively studied by earlier neutrino telescopes, this analysis served as an important benchmark for the performance of the newly installed hardware. Additionally, as the first analysis, it served as a platform for the development of new muon reconstruction algorithms, and a test of the new array simulation and techniques for modeling the ice properties.

 \begin{figure}[htbp]
\begin{center}
\includegraphics[width=3.0in]{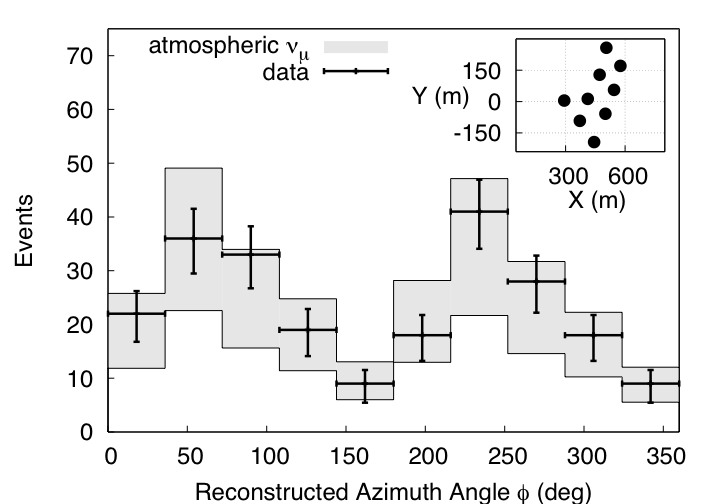}\includegraphics[width=3.0in]{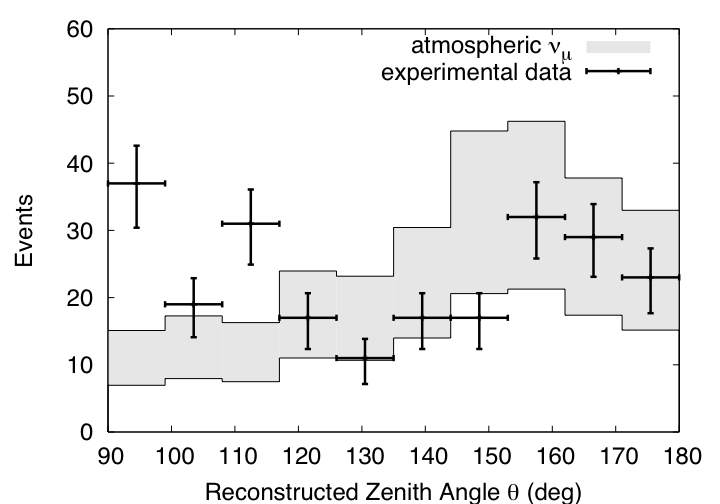}
\caption{The distribution of atmospheric neutrino candidates in azimuth (left) and zenith (right).  The shaded area is the range, including systematic uncertainties, of the expected signal flux.  The inset on the left shows the configuration of the nine string detector.  The acceptance for muon tracks peaks along the long axis of the array.~\cite{ic_atmnu}}
\label{fig:ic_atmnu}
\end{center}
\end{figure}
For this analysis, 234 candidates were selected, in agreement with the  $211\pm76.1 {\rm (syst.)} \pm 14.5 {\rm (stat.)}$ expected from atmospheric neutrinos.  
 
Figure~\ref{fig:ic_atmnu} shows the distribution in azimuth (left) and zenith (right) where the gray bands show the systematic uncertainty. As in the AMANDA analysis, the dominant systematic uncertainty stems from the theoretical uncertainty on the atmospheric flux normalization.  Above a zenith angle of 120 (with 180 degrees corresponding to a vertical upgoing event), the data agree well with expectation.   The excess between 120 degrees and the horizon is consistent with contamination from poorly reconstructed atmospheric muons.  
The azimuthal distribution contains two distinct peaks, an artifact of the rectangular geometry of the nine string detector (see inset, Figure~\ref{fig:ic_atmnu}),  which biases the candidate selection toward events traversing the long axis of the detector.  Since the completion of of this analysis, the acceptance of the detector has become more uniform with the addition of new strings which make the array more symmetric.

\subsection{Point source search}

The observation of extraterrestrial point sources of high energy neutrinos is a major goal of the current generation of neutrino telescopes.  Since neutrinos point back to their source, a point source would appear as a localized  excess over the distribution of atmospheric neutrinos.  For these analyses, the pointing resolution of the detector is critical, since the point spread function of the detector will determine the distribution of events from a single source, while the atmospheric neutrino background will be evenly distributed as a function of zenith angle.  The event selection for these analyses thus focusses on the selection of well measured tracks. In most analyses, the selection is also biassed toward higher energy events since the atmospheric spectrum is steeply falling with a spectral index of $\gamma=3.7$, while that of a point source would be harder with a spectral index between $\gamma=2$ and $\gamma=3$.

\begin{figure}[htbp]
\begin{center}
\includegraphics[width=5.5in]{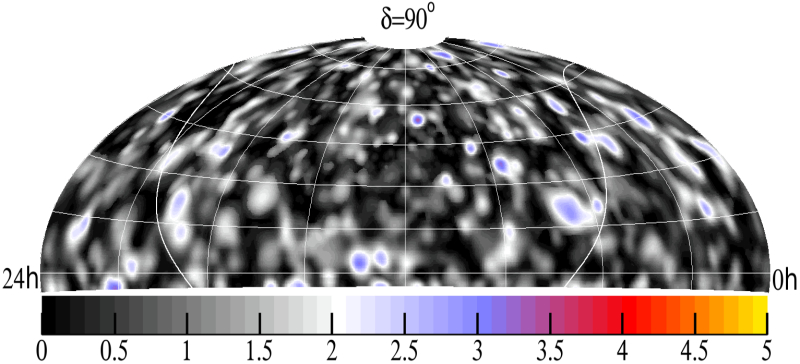}
\caption{Significance map from the AMANDA-II point source search using 6595 upward going candidates collected during 3.8 years of livetime from the years 2000-2006.  The regions with positive significance show an excess of events.  The solid lines show the galactic plane.~\cite{amanda_ps_new}}
\label{fig:am_ps}
\end{center}
\end{figure}

The most sensitive point search results to date have come from AMANDA, with the most recent published result using 3.8 years of livetime in the AMANDA-II detector.  While muon neutrino signatures are the mainstay of point source analyses, given their excellent pointing resolution, this analysis achieved an improvement in sensitivity by considering the contribution of tau neutrino interactions to the observed muon flux from the decay of the secondary tau via $\tau^\pm \rightarrow \mu^\pm + \bar{\nu}_\mu (\mu_\nu) + \nu_\tau (\bar{\nu}_\tau)$.   A previous AMANDA point source analysis calculated a 10-16\% improvement by using the tau neutrino contribution to the muon flux~\cite{am_ps}. The 6595 events selected were checked for consistency with the point source hypothesis via three methods.  First, the directions of the candidates were compared to the directions of 26 objects known to be sources of high energy gamma rays.  As outlined in Section 1.1, sites of hadronic acceleration are expected to produce gamma rays and neutrinos with correlated rates.  No significant excesses were found.  The largest clustering came from the directions of the Geminga ($\approx 2.6 \sigma$), which has a 20\% chance of random occurrence for one of 26 sources, and flux upper limits were set assuming a spectral index of $\gamma=2$.   Next, since no single source of the 26 individually tested showed a significant excess above background, six of the most significant sources from the  Milagro gamma ray telescope  were ``stacked''  to test their cumulative significance and again, no evidence of an astrophysical neutrino flux was found.  Finally, a scan of the Northern Hemisphere was used to search for a significant clustering of events (see Figure~\ref{fig:am_ps}).  The ``hottest'' spot showed a fluctuation of 3.38 $\sigma$, which would be expected in 95\% of random background trials and is therefore not significant.  
This analysis set a flux upper limit of $E^2\Phi^0_{\nu_\mu+\nu_\tau}=5.2 \times 10^{-11} {\rm TeV} {\rm cm}^{-2} {\rm s}^{-1}$ over the energy range 1.9 TeV to 2.5 PeV~\cite{amanda_ps_new}.

 \begin{figure}[htbp]
\begin{center}
\includegraphics[width=6.0in]{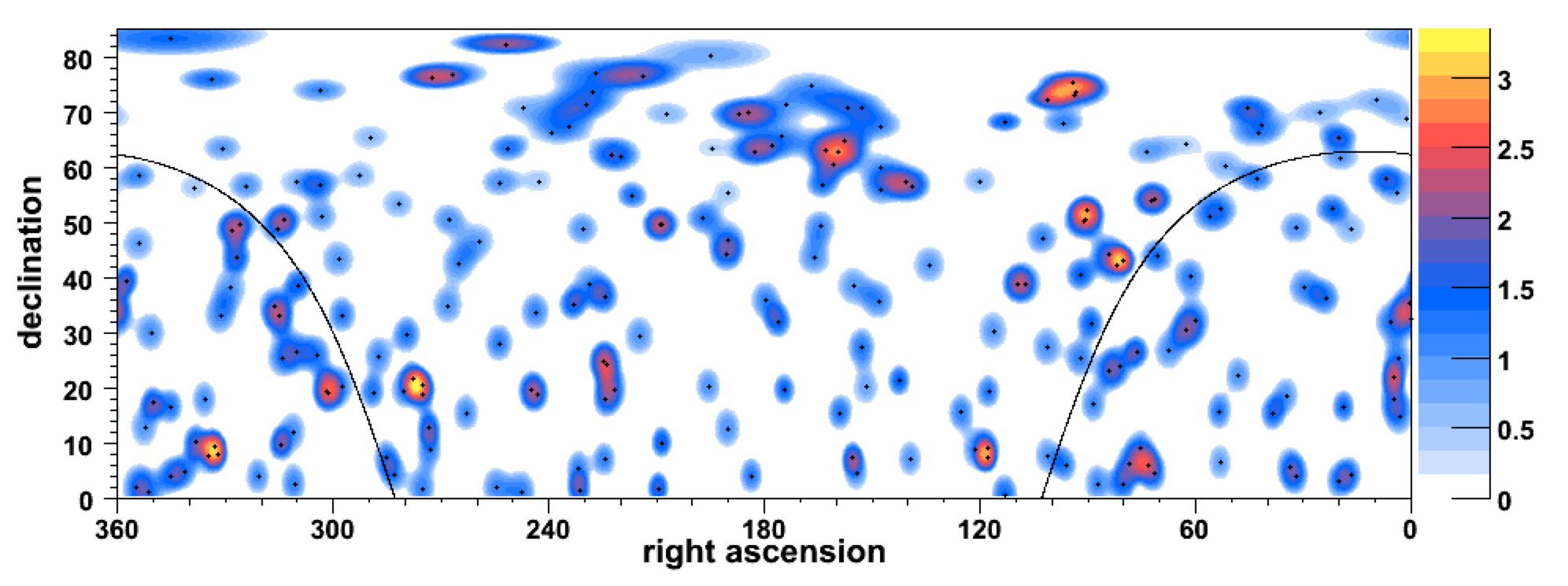}
\caption{Sky map from the IceCube 9 string point source search.  The black crosses show the reconstructed event directions and the shading shows the deviation from background in $\sigma$'s.~\cite{ic_ps}}
\label{fig:ic_ps}
\end{center}
\end{figure}

A point search search has also been performed using the same 2006 IceCube data sample that was used in the atmospheric neutrino analysis~\cite{ic_ps} .
Although the configuration of the detector is far from optimal, the point source sensitivity of the nine string
array is comparable to an equivalent livetime in the AMANDA-II detector.  Part of this gain is the result of the introduction of an unbinned likelihood technique~\cite{bins}, which improved the sensitivity 10\% over a binned search. 
The result is shown in Figure~\ref{fig:ic_ps}.  The maximum deviation from background expectation is 3.5 $\sigma$, which is not significant, since such a deviation has a 60\% chance of occurring in random background trials.  The likelihood was also calculated for a list of 26 preselected candidate sources.  These included some of the sources previously used in the AMANDA analysis, as well as some new sources added in response to a new analysis of the northern sky by the Milagro gamma ray telescope.  Of the sources in the updated catalogue, the most significant excess, $1.77 \ \sigma$, was in the direction of the Crab Nebula, and is also consistent with background fluctuations.

\begin{figure}
\begin{center}
\includegraphics[width=4.5in]{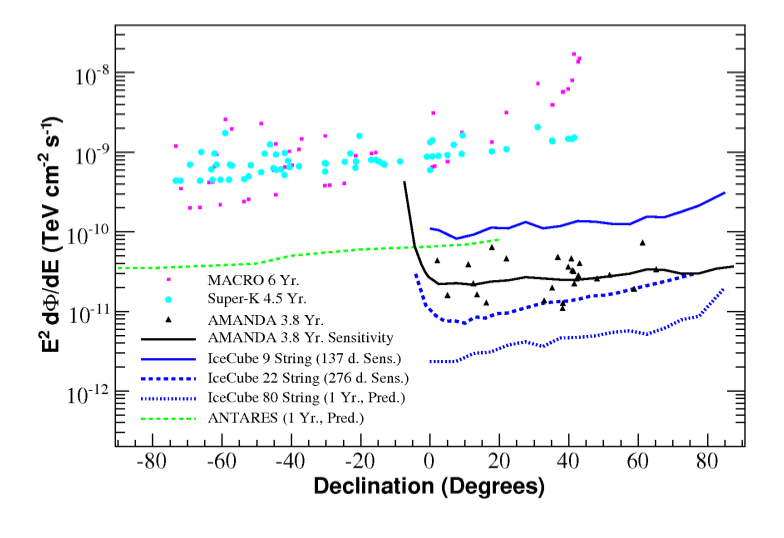}
\caption{The sensitivity of past and present neutrino observatories~\cite{amanda_ps_new,ant_ps}.  See text for details.}
\label{fig:ant_ps}
\end{center}
\end{figure}

Since the data used in this first IceCube point source analysis was taken, the ANTARES detector been completed, and the number of deployed IceCube strings has grown to 40.  The projected flux sensitivity of ANTARES to point sources as a function of declination is shown by the green dashed line in Figure~\ref{fig:ant_ps}.  Because of its excellent angular resolution, one year of data from the ANTARES detector is expected to approach the sensitivity of the 7 year AMANDA analysis, shown in black triangles on the same figure.  Note that the field of view of ANTARES overlaps AMANDA by about $1.5\pi$ steradians, allowing cross checks between the two instruments.  Also shown in the same Figure are point source limits from MACRO~\cite{macro}, a neutrino detector located in the deep underground laboratory at Gran Sasso, Italy, and Super-K.  From ANTARES' vantage point in the Northern hemisphere, it will therefore provide the most sensitive scan of the Southern sky.  A single year of IceCube data  taken with the full 80 strings of IceCube is expected to have a point source sensitivity in $E^2 \Phi$ of $5 \times 10^{-9} \ {\rm GeV}\,  {\rm cm}^{-2} \, {\rm s}^{-1} ({\rm sr}^{-1})$ averaged over the Northern sky, which is already significantly more sensitive than the AMANDA 2000-2006 analysis.  This is shown as the blue dotted line in Figure~\ref{fig:ant_ps}.

\subsection{Coincident observations of gamma ray bursts and other transient sources}

\begin{figure}[htbp]
\begin{center}
\includegraphics[width=4.0in]{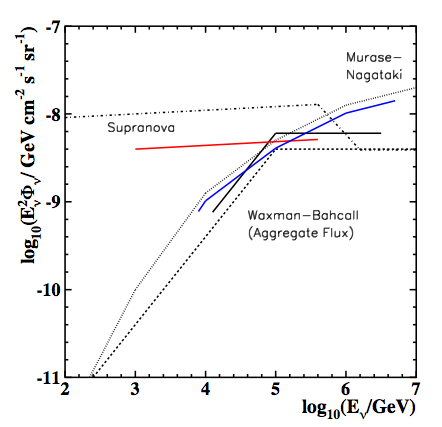}
\caption{The AMANDA limits on the muon neutrino flux from gamma ray bursts~\cite{am_grb_mu}. The fluxes predicted by the Waxman-Bahcall model~\cite{wb_grb} (dashed line), the Murase-Nagataki model~\cite{mn_grb} (dotted line), and the Razzaque model~\cite{raz_grb} (dash-dotted line) are shown for comparison.}
\label{fig:am_grb_mu}
\end{center}
\end{figure}

Because of their extreme energy, gamma ray  bursts are one of the most promising candidates for high energy cosmic ray acceleration.  By correlating neutrino observations to the observation of transient events by either ground or space based gamma ray telescopes, the statistical power of any point source observation is greatly increased, since the direction of the event and the time at which it occurred will be known.  In the past, AMANDA has performed searches for neutrinos arriving in coincidence with bursts observed by  the BATSE and IPN3 satellites.   A recently completed AMANDA analysis of 400 bursts for coincident muon neutrinos yielded no candidates.  This allowed a flux limit  to be set that was 1.3 times above  the flux of neutrinos from gamma ray bursts predicted by Waxman and Bahcall~\cite{wb_grb}  as shown in Figure~\ref{fig:am_grb_mu}.  The figure also shows the flux predictions and limits for other gamma ray burst models.  The limit on the model of Razzaque is already a factor of two below the most optimistic flux predictions of that model. The recent launch of the Swift satellite~\cite{swift} and the eminent launch of the GLAST mission~\cite{glast}, will provide an unprecedented number of bursts for study in the coming years.  Since the AMANDA detector was already tantalizingly close to the sensitivity needed to address these models, it has been estimated that the observation of only 70 bursts with the IceCube detector would be required to exclude these models at the 3 $\sigma$ confidence level if no candidates are observed~\cite{ic_grb}.

Individual events of special interest can be analyzed as well.  On December 27, 2004, gamma-ray satellites recorded the brightest transient ever observed, with most of the energy concentrated in the initial spike which lasted ~1 s.  The source was the Soft Gamma-ray Repeater 1806-20, thought to be a magnetar.  Several scenarios~\cite{flare1,flare2} have been put forward  to explain the flare that invoke a baryonic fireball and thus the possibility of high energy neutrino production. Although the source was in the Southern hemisphere, the precise knowledge of the time, location, and duration of the event allowed AMANDA to do a nearly background-free search.  Additionally, its Southern hemisphere location allowed limits on TeV gamma rays via downgoing muons produced by muon photoproduction in the atmosphere. In fact, this analysis was more sensitive to high energy gamma rays than neutrinos, since the overburden for neutrinos from the Southern sky neutrinos is limited.  No events were observed in 5.8 degree window around the source within 1.5 s of the flare, ruling out gamma ray spectral indices harder than $\gamma=1.5$~\cite{am_flare}.  

\subsection{Diffuse spectra of neutrinos}

\begin{figure}[b]
\begin{center}
\includegraphics[width=4.5in]{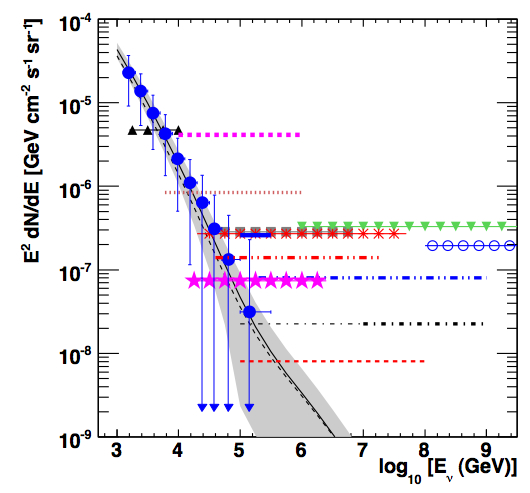}
\caption{The limits on the diffuse flux of neutrinos.  Currently, the most stringent limit is from AMANDA II (purple stars).  The sensitivity of 1 year of data with the full IceCube array is given given by the red dotted line.  This assumes an energy spectrum that falls as $E^{-2}$~\cite{ic_diffuse}.  See text for details of the other limits shown.}
\label{fig:ic_diffuse}
\end{center}
\end{figure}

Combining events from isotropically distributed sources can yield evidence of an extraterrestrial neutrino flux, even if the individual sources are not identified.  Such events would be apparent as an excess over the measured atmospheric neutrino flux (blue solid dots in Figure~\ref{fig:ic_diffuse}), and since cosmic sources are expected to have a harder spectrum, energy correlated observables can be used to select candidate events.  The most sensitive diffuse astrophysical neutrino limit to date comes from AMANDA, which set a 90\% confidence level upper limit on the flux of muon neutrinos of $E^2\Phi < 7.4 \times 10^{-8}$ GeV ${\rm cm}^2$ ${\rm s}^{-1}$ ${\rm sr}^{-1}$ in the energy range of 16 TeV to 2.5 PeV assuming an $E^{-2}$ spectrum (see Figure~\ref{fig:ic_diffuse}  purple stars) using 807 days of data collected between 2000 and 2003.  This limit was nearly an order of magnitude stronger than a previous limit from AMANDA B-10 (red dotted line).  Since directional information isn't important in diffuse searches all three neutrino flavors can be used to set an upper limit on the flux.   The red asterisks show the flux limits from such a search from Baikal.  An all-flavor search from AMANDA B-10 is slightly weaker, but extends to higher energies (green triangles).   Both of these flux limits are still well above the Waxman-Bahcall prediction discussed in Section 1.3, shown by the heavy black dash-dotted line in the same figure.  

\begin{figure}[b]
\begin{center}
\includegraphics[width=4.0in]{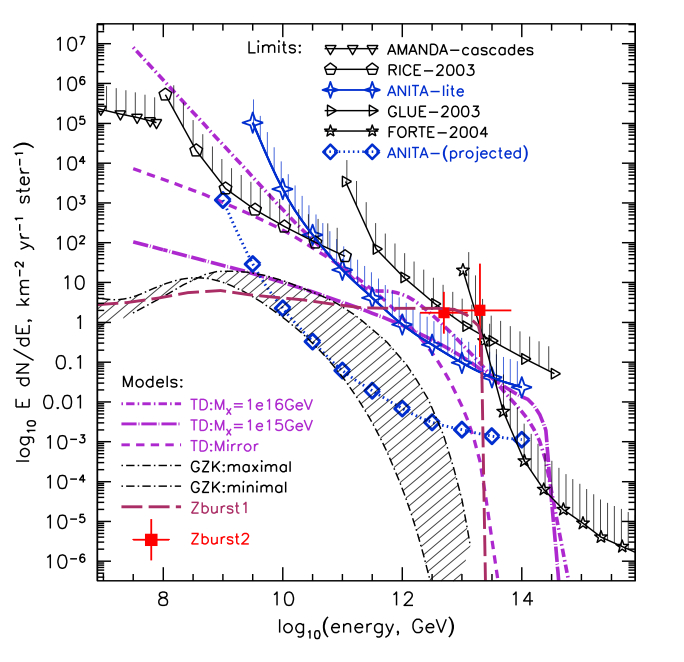}
\caption{The limits on the diffuse flux of neutrinos at very high energies.~\cite{anita} See text.}
\label{fig:anitalite}
\end{center}
\end{figure}

The search for ultra high energy neutrinos is of particular interest because they should be produced in the interaction of cosmic rays on the cosmic microwave background, producing a guaranteed flux.  In addition, they too could be produced as a by product of hadronic acceleration AGNs.  Such a search could also be sensitive to exotic phenomena.  However, this requires a dedicated search for very high energy tracks emanating from the horizon because the Earth is opaque to neutrinos above $10^7$ GeV.
The most stringent flux limits up to an energy of $10^9$ GeV come from an analysis of 571 live days taken with AMANDA between 2000 and 2002 (blue dash-dotted line in Figure~\ref{fig:ic_diffuse}).   RICE has set flux limits that extend to higher energies still (blue open circles).  Figure~\ref{fig:anitalite} shows the expected GZK flux and a summary of existing and projected limits at ultra high energies. Anita-lite (blue stars in Figure~\ref{fig:anitalite}) was a proof of concept experiment for ANITA flown in 2004.  The results of the 2006-2007 ANITA flight will be available shortly.  As shown by the blue diamonds, ANITA is expected to have sufficient sensitivity to detect a GZK neutrino.  If it succeeds, it will be the first observation of neutrinos via the Askaryan effect.  In addition, searches for an ultra high energy flux from extensive air shower arrays are underway, and the first results from Auger are now available.~\cite{auger_nu}
Using data collected between January 2004 and August 2007, Auger obtains a limit of 130 eV ${\rm sr^{-1} \, s^{-1} \, cm^{-2}}$ in the energy range of 17.3 to 19.3 $\log_{10}$ (E/eV), assuming an $E^{-2}$ spectrum.  Since the Auger detector was under construction during this time, this data only corresponds to an equivalent  livetime of about one year of integrated data with the completed array.  This is the most sensitive limit to date in the EeV energy range.


\begin{figure}[b]
\begin{center}
\includegraphics[width=4.0in]{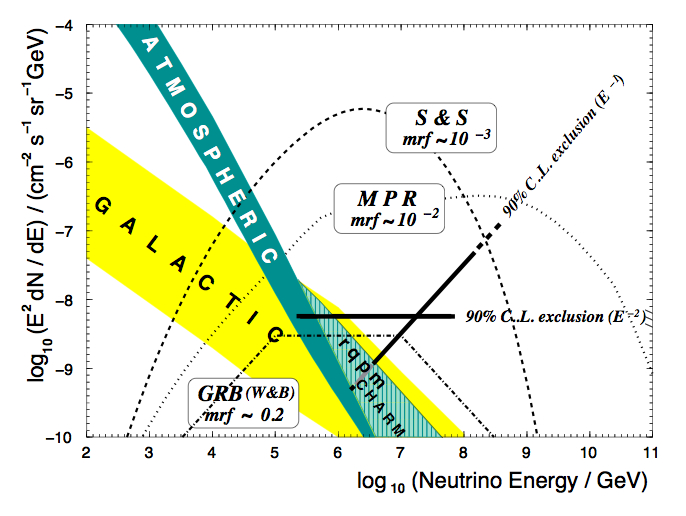}
\caption{The sensitivity of IceCube to various neutrino flux calculations~\cite{ic_sens} . The solid lines show the 90\% C.L. upper imit after three years of data taking.  The dashed line indicates the Stecker and Salamon~\cite{ss} model for photo-hadronic 
interactions in AGN cores. The dottedl ine corresponds to the Mannheim, 
Protheroe and Rachen upper bound discussed in Section 1.1. Also shown is the GRB model by Waxman and Bah- 
call (dash-dotted line), where the model rejection factor was calculated for 500 bursts.}
\label{fig:ic_model}
\end{center}
\end{figure}

As discussed in Section 1.3, the sensitivity of an array to a diffuse glow of neutrinos is a useful benchmark since the expected integrated flux is well motivated by cosmic ray observations.
One year of data from ANTARES has an expected sensitivity to a diffuse flux of $0.9 \times 10 ^{-7}$ ${\rm GeV} \ {\rm cm^{-2}}  {\rm s}^{-1} {\rm sr}^{-1}$.     A preliminary study of the sensitivity of the 137 days of data taken with the only the first nine IceCube strings array is shown by the red dash-dotted line in Figure~\ref{fig:ic_diffuse}.  The analysis was based on muon neutrino signatures, and it shows that the instantaneous sensitivity (the sensitivity per unit time) of IceCube-9 is already a factor of 2 higher than the instantaneous sensitivity of the AMANDA detector.  One year of data taken with the full 80 strings of IceCube will have a sensitivity to fluxes of $7 \times 10^{-9}$ ${\rm GeV} \ {\rm cm^{-2}}  {\rm s}^{-1} {\rm sr}^{-1}$ (red dashed line)~\cite{ic_sens}.  This shows that one year of data with the full IceCube array is sensitive to diffuse fluxes well below the Waxman-Bahcall limit assuming an $E^{-2}$ spectrum.   The expected sensitivity of IceCube to other neutrino flux predictions (see Section 1.1) after three years of data taking are shown in Figure~\ref{fig:ic_model}.  The sensitivities are quoted in terms of the model rejection factor (mrf), meaning that IceCube will be sensitive to spectra of the same shape, but with fluxes smaller by the amount of the mrf. 
Neutrino astronomy is entering an exciting and promising new era in sensitivity with the ANTARES and IceCube construction nearing completion.

\subsection{Particle physics}

As recounted in the introduction, the most significant discoveries from neutrino telescopes to date have been in the field of particle physics.  As the world's largest particle physics detectors, these instruments have unique capabilities as long baseline neutrino experiments and in the search for very rare or weakly interacting particles.  Two phenomena on which neutrino telescopes have consistently set competitive limits are magnetic monopoles and weakly interacting massive particles (WIMPs).  If they exist, the flux of monopoles must be very low since they would take energy from the galactic magnetic field, resulting in a limit called the Parker Bound.  Such an object would manifest itself in a Cherenkov detector as a slow moving track producing 8000 times the light expected from a muon.  If  unknown massive but weakly interacting particles such as a neutralino existed, they would be trapped in the gravity well of the Sun and the Earth.  Subsequent WIMP annihilations would produce neutrinos at the end of the decay chain, making the Sun and the center of the Earth an apparent source of neutrinos.

\begin{figure}[htbp]
\begin{center}
\includegraphics[width=4in]{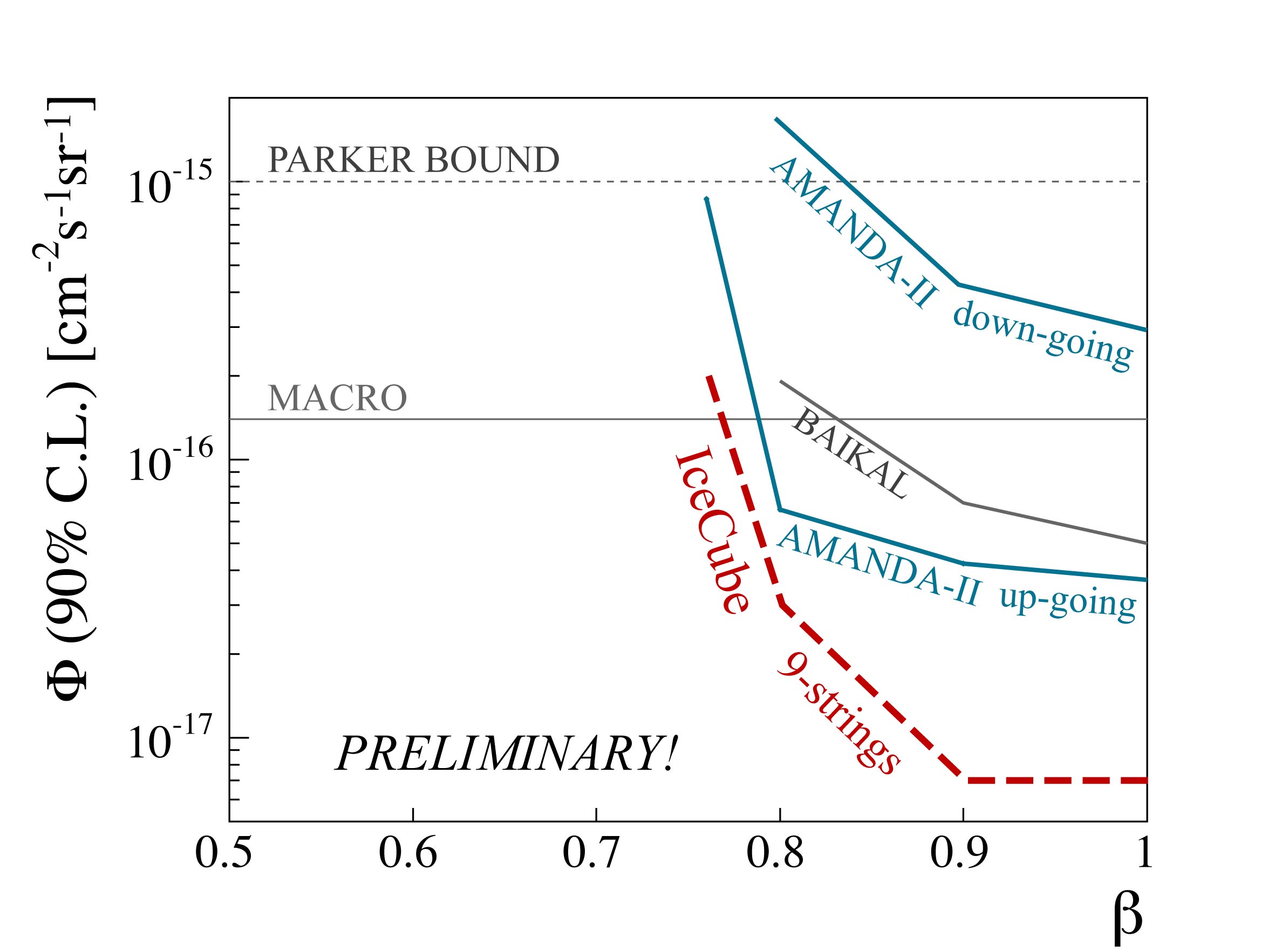}
\caption{90\% confidence level flux limits on monopoles as a function of monopole speed~\cite{mono}.  The AMANDA-II limits come from a recent analysis of one year of data.  The limits  from MACRO are also shown~\cite{macro_mono}. The projected IceCube limit shown is from a study of the sensitivity of the 9 string detector.}
\label{fig:mono}
\end{center}
\end{figure}

A recent analysis of one year of AMANDA data resulted in the most stringent limit on monopoles to date  for high values of $\beta$ (Figure~\ref{fig:mono}), eclipsing previous limits set by Baikal (black line)~\cite{baikal_mono} and MACRO~\cite{macro_mono}.  Due to their extreme brightness, even downgoing monopoles may be sought with minimal background.  An analysis of monte carlo events for the IceCube detector in its nine string configuration shows that, even with only 137 days of livetime, the sensitivity of this small portion of the full IceCube detector will already supercede these limits.

\begin{figure}[htbp]
\begin{center}
\includegraphics[width=4in]{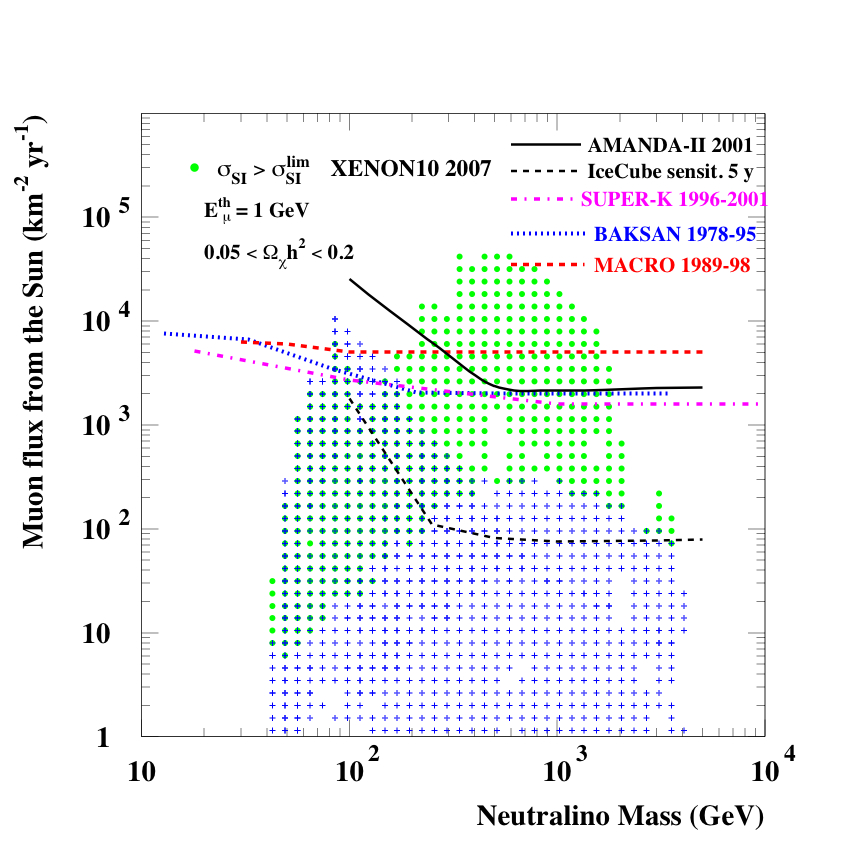}
\includegraphics[width=4in]{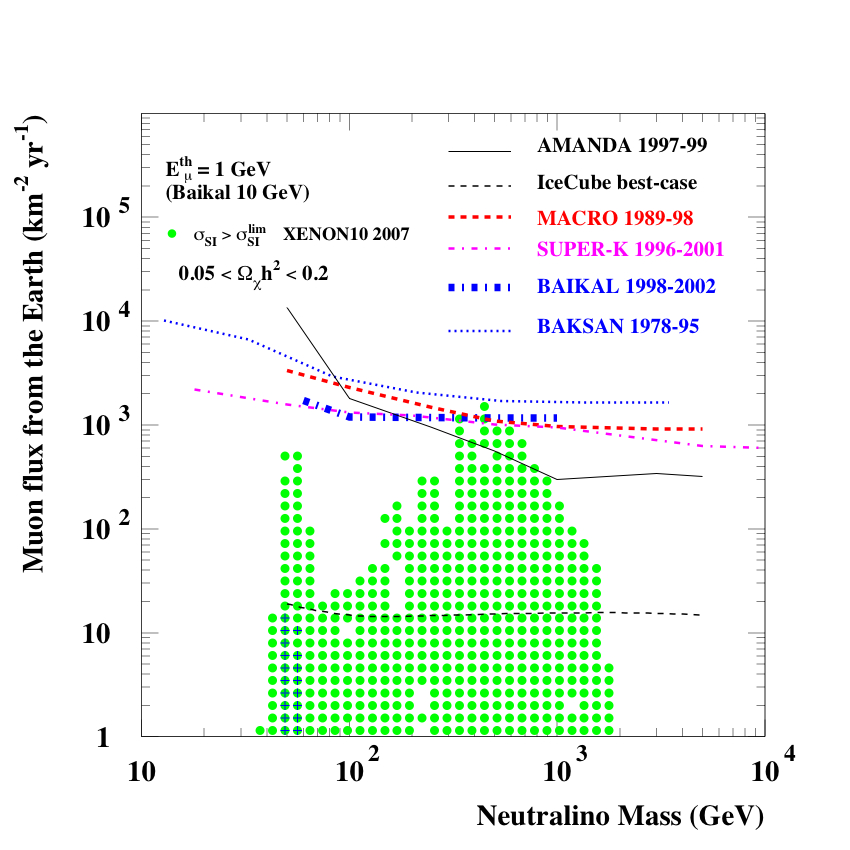}
\caption{Limits on WIMPs emanating from the Sun (top)~\cite{wimp_am2} and the Earth (bottom)~\cite{wimp_am} from indirect detection by neutrino experiments (lines) and direct detection (green dots) by XENON10~\cite{xenon10} Dark Matter Experiment located at the Gran Sasso Laboratory.  The blue crosses show the phase space favored by cosmologically relevant MSSM models.}
\label{fig:wimps}
\end{center}
\end{figure}

A summary of  searches for neutralino WIMPS invoked in cosmologically relevant minimal supersymmetric models is shown in Figure~\ref{fig:wimps}, where the top plot shows the limits for WIMPs annihilating in the Sun and and the bottom shows the limits for WIMPs annihilating at the center of the Earth.   For the higher energy neutrino telescopes, the theoretically favored masses are at the low end of the detection sensitivity, making the limits from smaller telescopes such as Baikal (blue dash-dotted line) and Super-K (purple dash dotted line) still very competitive.  Solar WIMPs will produce signatures that traverse the detectors located at far North and South latitudes horizontally.  AMANDA B-10 and Baikal had a strongly vertical detector configuration which therefore limited their sensitivity to solar WIMPs.

The expected sensitivity of IceCube is indicated in the same plot by the black dashed lines.  Because IceCube's phototubes are spaced more densely along its vertical axis than its horizontal axis, it will be much more sensitive to Earth WIMPs than solar WIMPs.  
The sensitivity of IceCube to solar WIMPs is further diminished by the fact that the Sun sinks maximally 23 degrees below the horizon at the South Pole, and for this reason, the Earth cannot be used as effectively to filter the background from atmospheric muons in solar WIMP searches.  
The sensitivity to horizontal tracks and low energy phenomena can be improved by using  the AMANDA array which has now been surrounded by the IceCube strings.  By using the IceCube strings as a veto, low energy tracks may be distinguished from background, selecting events that start and stop within the instrumented area.~\cite{WIMP}   Once completed, the ``deep core'' discussed in Section 2.1.2, will greatly extend the sensitivity of IceCube to lower energy phenomena, including WIMPS.

\section{Summary}

It is truly an exciting time in particle astrophysics. Building on the long history of cosmic-ray and gamma-ray measurements, new experiments are coming on line that significantly extend our ability to observe the highest energy astrophysical objects, giving us new insight into the mechanisms at work in these objects and perhaps into the old question of the source of the cosmic rays. These new instruments include high energy cosmic ray detectors, gamma-ray detectors on the surface and in space, and neutrino telescopes with sensitivities that are of astrophysical interest. 

Neutrino telescopes have a unique and complementary role to play in this new era of particle astronomy with their ability to look for distant sources at energies not accessible to gamma-ray instruments, and because neutrinos, as new messengers, can provide unique information on the acceleration mechanisms at work in these high energy astrophysical sources. The next generation neutrino observatories will give us an unprecedented sensitivity for sources in both the northern and southern skies. With detectors on the scale of a cubic kilometer operating at South Pole in the near future, and planned for the Mediterranean soon after, we are optimistic that more than thirty years since they were first conceived, we will finally open this new window on the high energy universe.

\ack
The author wises to thank John Carr for his valuable input on the status of the ANTARES array, and Greg Sullivan and Jordan Goodman for their helpful comments on this manuscript.
This work was supported by the National Science Foundation  through grant No. PHY-0502709.

\end{document}